\documentclass[reprint, aps,prd,twocolumn,showpacs,showkeys, 10pt, longbibliography, nolinenumbers, floatfix]{revtex4-1}

\usepackage{amsmath}
\usepackage{graphicx}
\usepackage{float}
\usepackage{dcolumn}
\usepackage{multirow}
\usepackage{natbib}
\usepackage[colorlinks = true,linkcolor = blue,urlcolor  = blue,citecolor = blue,anchorcolor = blue]{hyperref}

\begin{document}

\title{Speed of sound constraints on maximally rotating neutron stars}

\author{Ch. Margaritis}
\email{chmargar@auth.gr}

\author{P.S. Koliogiannis}
\email{pkoliogi@physics.auth.gr}

\author{Ch.C. Moustakidis}
\email{moustaki@auth.gr}

\affiliation{Department of Theoretical Physics, Aristotle University of Thessaloniki, 54124 Thessaloniki, Greece}

\begin{abstract}
The observation of maximally rotating neutron stars (in comparison to nonrotating ones) may provide more information on the behavior of nuclear matter at high densities. We provide a theoretical treatment concerning the effects of the upper bound of the sound speed in dense matter on the bulk properties of maximally rotating (at mass-shedding limit) neutron stars. In particular, we consider two upper bounds for the speed of sound, $v_s = c$ and $v_s = c/\sqrt{3}$, and the one provided by the relativistic kinetic theory. We investigate to what extent the possible predicted (from various theories and conjectures) upper bounds on the speed of sound constrain the ones of various key quantities, including the maximum mass and the corresponding radius, Keplerian frequency, Kerr parameter and moment of inertia. We mainly focus on the lower proposed limit, $v_{s}=c/\sqrt{3}$, and we explore  in which mass region a rotating neutron star collapses to a black hole. In any case, useful relations of the mentioned bulk properties with the transition density are derived and compared with the corresponding nonrotating cases. We concluded that the proposed limit $v_{s}=c/\sqrt{3}$ leads to dramatic decrease on the values of the maximum mass, Kerr parameter and moment of inertia preventing a neutron star to reach values which derived with the consideration of realistic equations of state or from other constraints. Possible measurements of the Kerr parameter and moment of inertia would shed light on these issues and help to reveal the speed of sound bound in dense matter.  


\end{abstract}

\maketitle


\section{Introduction}\label{sec:1}
Neutron stars are the most suitable extraterrestrial laboratories to study the properties of dense matter under extreme conditions~\cite{Shapiro-1983,Glendenning-2000,Haensel-2007,Hartle-1978}. Their rapid rotation (due to very strong compactness) exhibits high spin frequency and the mass velocity at the surface may be close to the speed of light~\cite{Friedman-2013,Paschalidis-2017}. In any case, it is a common belief that the study (observational and theoretical) of the maximally rotating neutron stars may offer rich information on the properties of dense nuclear matter. In particular, the derivation of the Keplerian frequency (the frequency at which a rotating star would shed matter at its equator) has a twofold meaning. In Newtonian theory it takes a simple form and comes from the balance between gravitational and centrifugal forces. In General Relativity it is expressed as a self-consistency condition that must be satisfied by the solution to Einstein's  equations. It should be noted that there are no analytical solutions of these equations for rotating neutron stars and consequently, only numerical  estimations exist for  the Keplerian frequency.\\
\indent The strong-interaction between many bodies has been the main mechanism for the theoretical construction of the equation of state (EoS). The main assumption is that the speed of sound in an EoS cannot exceed the speed of light because of the causality. But a question rises: ``\textit{Is the speed of light the upper bound of the speed of sound in dense matter?}". In fact, Hartle in Ref.~\cite{Hartle-1978} pointed out that causality is not enough to constrain the high density part of the EoS. In recent years, non-relativistic and weakly coupled theories have provided us with the $c/\sqrt{3}$ bound through classes of strongly coupled theories with gravity duals~\cite{Bedaque-2015}. Bedaque and Steiner pointed out that the existence of neutron stars with masses about two solar masses, combined with the knowledge of the EoS of hadronic matter at low densities, is not consistent with this bound~\cite{Bedaque-2015}.\\
\indent The value of the speed of sound in dense matter is still an open problem. Recently, the effects of the upper bound of the sound speed in neutron star properties have been studied extensively~\cite{Glendenning-2000,PhysRevC.95.045801,Friedman-1988,Lattimer-1990,Koranda-1997,Haensel-2009,Rhoades-1974,Chamel-2013a,PhysRevD.95.083014,Silva_2017,Alsing-2018,Podkowka-2018,Miller-2019,xia2019sound,PhysRevC.99.035803,BAIOTTI2019103714,PhysRevD.100.114003}. However, the majority of the studies where focused on nonrotating or slow-rotating neutron stars. Haensel and Zdunik~\cite{Haensel-1989} were the ones who first discussed the causality as a bound on the maximum angular velocity of uniformly rotating neutron stars. Later on, Glendenning estimated the effects of the speed of sound on the period of gravitationally bounded stars. In particular, he employed a nonrotating model and estimated the period using an empirical formula connecting the period with the bulk properties of nonrotating neutron stars~\cite{Glendenning-1992}. Lattimer {\it et al.}~\cite{Lattimer-1990} and Koranda {\it et al.}~\cite{Koranda-1997} have also studied the upper limits set by causality on the rotation and mass of uniformly rotating relativistic stars. It is worth pointing out that, while in Ref.~\cite{Koranda-1997} the upper limit set  by causality, which is $v_s/c=1$, was the only limit that has been considered, in Ref.~\cite{Glendenning-1992} the parametrization of the speed of sound was took effect in case of quark stars.\\
\indent In the present work, we employ in addition two upper bounds, the $v_s/c=1/\sqrt{3}$ and the one originated from the relativistic kinetic theory~\cite{Israel-1976,Israel-1979,Hiscock-1983,Olson-1989,Olson-1990,Olson-2000}. In the latter one, the sound speed is introduced by the corresponding theory as a self-consistent way and not artificial. The main motivation of the present work is to investigate the possibility to provide some universal constraints on the bulk properties of maximally rotating neutron stars including the maximum mass and the corresponding radius, Keplerian frequency, Kerr parameter and moment of inertia. All these quantities are sensitive on the high density part of the EoS and we expect that the imposed constraints on the stiffness of the EoS may lead to important constraints on the dense nuclear matter. The EoS that is used is predicted by the Momentum-Dependent-Interaction (MDI) model in correlation with data from Akmal {\it et al.}~\cite{Akmal-1998} and predicts the current observed maximum neutron star masses ($1.908\pm 0.016M_{\odot}$~\cite{Arzoumanian-2018}, $2.01\pm 0.04M_{\odot}$~\cite{Antoniadis-2013}, $2.14^{+0.10}_{-0.09}M_{\odot}$~\cite{Cromartie-2019}, $2.27^{+0.17}_{-0.15}M_{\odot}$~\cite{Linares-2018}) (for more details see Ref.~\cite{Koliogiannis-2019}). We also study the constant rest mass sequences of a neutron star in order to provide constraints relative to its collapse to a black hole~\cite{Koliogiannis-2019}. Finally, we provide a detailed study about the connection between the minimum period of a rotating neutron star and the maximum neutron star mass of a nonrotating one.\\
\indent We consider that the observational measurements, in correlation with the theoretical studies of maximally rotating neutron stars, will help significantly to gain rich information about the properties of dense nuclear matter. For example, the remnant of the GW170817 merger may lead to differentially rotating neutron star close to mass-shedding limit (Keplerian frequency)~\cite{Lattimer-2019}. In this case, the detection of the gravitational waves from this source may provide useful information for the dense matter properties~\cite{Tews-2019}. In fact, the bound $v_s = c/\sqrt{3}$ ``controls" the stiffness of the EoS at high densities (since in most EoSs the speed of sound exceeds the value $v_s = c/\sqrt{3}$ even for low values of the density) and apparently, affects appreciably the bulk properties of both nonrotating and rotating neutron stars.\\
\indent The article is organized as follows: In Section~\ref{sec:2} we present the speed of sound bounds and the nuclear equation of state model. In Section~\ref{sec:3} we briefly review the rotating configuration on neutron stars, while in Section~\ref{sec:4}, we introduce the effects of the speed of sound bounds on the bulk properties of neutron stars. In addition, a detailed study about the constant rest mass sequences and the minimum rotational period is provided. Finally, Section~\ref{sec:5} includes the relevant discussion as well as the conclusions of the present study.  

\section{Speed of sound bounds  and  maximum mass configuration} \label{sec:2}
We have constructed the maximum mass configuration (we consider that to a very good accuracy this configuration is identified with the maximum permitted rotational frequency (Keplerian velocity)~\cite{Friedman-2013,Paschalidis-2017})  by considering the following two structures for the neutron star EoS:
\begin{itemize}
\item[a)] {\it Maximum angular velocity for known low-density EoS}
\item[b)] {\it Maximum angular velocity from the relativistic kinetic theory}
\end{itemize}
In the first case, (a), the EoS is given through the ansatz~\cite{PhysRevC.95.045801}

\begin{eqnarray}
P({\cal E})&=&\left\{
\begin{array}{ll}
P_{\rm crust}({\cal E}), \quad  {\cal E} \leq {\cal E}_{\rm c-edge}  & \\
\\
P_{\rm NM}({\cal E}), \quad  {\cal E}_{\rm c-edge} \leq {\cal E} \leq {\cal E}_{\rm tr}  & \\
\\
\left(\frac{v_{\rm s}}{c}  \right)^2\left({\cal E}-{\cal E}_{\rm tr}  \right)+
P_{\rm NM}({\cal E}_{\rm tr}), \quad  {\cal E}_{\rm tr} \leq {\cal E}  . & \
\end{array}
\right.
\label{EOS-1}
\end{eqnarray}
where $P$ and ${\cal E}$ are the pressure and energy density, respectively, and $\mathcal{E}_{\rm tr}$ is the transition energy density. In the second case, (b), the maximally stiff EoS fulfills the following expression

\begin{equation}
\left(\frac{v_{\rm s}}{c}\right)^2= \frac{{\cal E}-P/3}{P+{\cal E}},
\label{equal-kin-4}
\end{equation}

\noindent and the EoS is given through the ansatz~\cite{Olson-2000,PhysRevC.95.045801}

\begin{eqnarray}
P(n)&=&\left\{
\begin{array}{ll}
P_{\rm crust}(n), \quad  n \leq n_{\rm c-edge} & \\
\\
P_{\rm NM}(n), \quad  n_{\rm c-edge} \leq n \leq n_{\rm tr} & \\
\\
{\cal C}_1n^{a_1}(a_1-1)+{\cal C}_2n^{a_2}(a_2-1)
, \quad   n_{\rm tr} \leq n  & \
\end{array}
\right.
\label{Kinetic-eos}
\end{eqnarray}

\noindent with 
\begin{equation}
\alpha_{1} = (1+\sqrt{13})/3, \quad \alpha_{2} = (1-\sqrt{13})/3,
\end{equation}
\begin{equation}
\mathcal{C}_{1} = \left(\frac{(2+\sqrt{13})\mathcal{E}(n_{\rm tr})+3P(n_{\rm tr})}{2\sqrt{13}}\right)n_{\rm tr}^{-\alpha_{1}},
\end{equation}
\begin{equation}
\mathcal{C}_{2} = -\left(\frac{(2-\sqrt{13})\mathcal{E}(n_{\rm tr})+3P(n_{\rm tr})}{2\sqrt{13}}\right)n_{\rm tr}^{-\alpha_{2}}.
\end{equation}
where $n$ is the baryonic density and $n_{\rm tr}$ is the corresponding transition density. In this case (b), the speed of sound is self-constrained in the range
\begin{equation}
\frac{1}{3} \leq \left(\frac{v_{\rm s}}{c}\right)^2 \leq 1.
\label{eq:sos_limits}
\end{equation}
In addition, Eq.\eqref{eq:sos_limits} shows the applicable range of each limit of the speed of sound which are studied in this paper.

\begin{table*}
	\squeezetable
	\caption{Parameters of the Eq.~\eqref{speed-matc-1} for the different transition densities and the various speed of sound bounds.}
	\begin{ruledtabular}
		\begin{tabular}{ccccccccccc}
			\multirow{2}{*}{Speed of sound bounds} & \multicolumn{2}{c}{$n_{\rm tr}=1.5n_{\rm s}$} & \multicolumn{2}{c}{$n_{\rm tr}=2n_{\rm s}$} & \multicolumn{2}{c}{$n_{\rm tr}=3n_{\rm s}$} & \multicolumn{2}{c}{$n_{\rm tr}=4n_{\rm s}$} & \multicolumn{2}{c}{$n_{\rm tr}=5n_{\rm s}$} \\
			& $c_{1}$ & $c_{2}$ & $c_{1}$ & $c_{2}$ & $c_{1}$ & $c_{2}$ & $c_{1}$ & $c_{2}$ & $c_{1}$ & $c_{2}$ \\
			\hline
			$c$ & 0.880 & 0.240 & 0.814 & 0.320 & 0.676 & 0.480 & 0.357 & 0.640 & 0.254 & 0.801 \\
			
			$c/\sqrt{3}$ & 0.214 & 0.240 & 0.147 & 0.320 & 0.009 & 0.480 & -- & -- & -- & -- \\
		\end{tabular}
	\end{ruledtabular}
	\label{tab:1}
\end{table*}

\indent To be more specific, as Eqs.~\eqref{EOS-1} and ~\eqref{Kinetic-eos} show, the EoS is divided in three regions. In region ${\cal E} \leq {\cal E}_{\rm c-edge}$, we used the equation of Feynman, Metropolis and Teller~\cite{Feynman-1949} and also of Baym, Bethe and Sutherland~\cite{Baym-1971} for the crust and low densities of neutron star. In the  intermediate region, ${\cal E}_{\rm c-edge} \leq {\cal E} \leq {\cal E}_{\rm tr}$, we employed a specific EoS based on the MDI model and data from Akmal {\it et al.}~\cite{Akmal-1998}, while for ${\cal E}_{\rm tr}\geq \mathcal{E}$, the EoS is maximally stiff with the speed of sound $\sqrt{\left(\partial P / \partial {\cal E}\right)}_{\rm S}$ fixed at two values, $c/\sqrt{3}$ and $c$. Although the energy densities below the ${\cal E}_{\rm c-edge}$ have negligible effects on the maximum mass configuration, we used them in calculations for the sake of completeness. The cases which took effect in this study were the ones where the fiducial transition density is $n_{\rm tr} = p_{\rm n} n_{\rm s}$, where $n_{s}$ is the saturation density of symmetric nuclear matter ($n_{\rm s}= 0.16$ $fm^{-3}$) and $p_{\rm n}$ takes the values $1.5,2,3,4,5$.\\
\indent In the specific case where $n_{\rm tr} =1.5n_{\rm s}$, when the speed of sound is equal to $c$, we have the EoS-maxstiff scenario; when the speed of sound is equal to $c/\sqrt{3}$, we have the EoS-minstiff scenario. However, since the transition density $n_{\rm tr}$ and the upper bound of the speed of sound are not fixed from first principles, the EoS remains unknown.\\ 
\indent In approach (a) the continuity on the  EoS is well ensured. This is in contrast to similar approach employed in Ref.~\cite{PhysRevD.88.083013} where a sharp phase transition is indicated. However, approach (a), owing to its artificial character,  does not ensure continuity in the speed of sound at the transition density. Therefore, we concentrated our study to treat the discontinuity in the speed of sound by enforcing a smooth phase transition. In particular, we avoided the discontinuities in the speed of sound appearing at the transition densities by employing a method presented in Ref.~\cite{Tews-2018}. We proceeded with the matching of the EoSs on the transition density  by considering that, above this value, the speed of sound is parametrized as follows (for more details see~\cite{Tews-2018})
\begin{equation}
\frac{v_{\rm s}}{c}=\left(a-c_1\exp\left[-\frac{(n-c_2}{w^2} \right]\right)^{1/2}, \quad a=1, 1/3
\label{speed-matc-1}
\end{equation}
where the parameters $c_1$, $c_2$, which are shown in Table~\ref{tab:1}, and $w=0.001$ are fit to the speed of sound and its derivative at $n_{\rm tr}$, and also to the demands $v_{\rm s}(n_{\rm tr})=[c, c/\sqrt{3}]$. Using Eq.~\eqref{speed-matc-1}, the EoS for $n \geq n_{tr}$ can be constructed with the help of the following recipe~\cite{Tews-2018}
\begin{equation}
{\cal E}_{i+1} = {\cal E}_i+\Delta {\cal E}, \quad P_{i+1} = P_i+\left(\frac{v_s}{c}(n_i)\right)^2\Delta {\cal E},
\label{eq:5}
\end{equation}
\begin{equation}
\Delta {\cal E} = \Delta n\left(\frac{{\cal E}_i+P_i}{n_i} \right),
\label{eq:6}
\end{equation} 
\begin{equation}
\Delta n = n_{i+1}-n_i.
\label{eq:7}
\end{equation}

Morover, in the approach (a) we studied the method where discontinuities are presented in the EoS and the one where continuity exhibits based on Eqs.~\eqref{speed-matc-1}-\eqref{eq:7}. The results are presented and discussed in Section~\ref{sec:4h}.\\
\indent In approach (b), while the continuity on the EoS is well ensured by definition (relativistic kinetic theory), the speed of sound exhibits an inevitable discontinuity at the transition density. In particular, a sharp phase transition in the speed of sound appeared in the case of the maximally stiff EoS. Although employ an approach similar to case (a) would treat the relevant discontinuity, these effects on the bulk properties of neutron stars at the maximum mass configuration are insignificant (as has been found from previous work and the present study). Thus, in case (b), the discontinuity on the speed of sound is taken into account without any approximation. 

\section{Rotating neutron stars} \label{sec:3}
It has been shown by Friedman {\it et al.}~\cite{Friedman-1988} that the turning-point method, which is leading to the points of secular instability, can also be used in the case of uniformly rotating neutron stars. With this consideration, in a constant angular momentum sequence, the turning-point of a sequence of configurations with increasing central density, separates the secular stable from unstable configuration and consequently, the condition
\begin{equation}
\frac{\partial M({\cal E}_{\rm c},J)}{\partial {\cal E}_{\rm c}}\mid_{\rm J={\rm constant}}=0,
\label{cond-1}
\end{equation}
where ${\cal E}_c$ is the energy density in the center of the neutron star and $J$ is the angular momentum, defines the possible maximum mass.\\
\indent The Keplerian angular momentum is obtained as a self-consistency condition in the solution of Einstein's equations for a rotating neutron star. In this case an approximate expression has been obtained according to~\cite{Glendenning-2000}
\begin{equation}
\Omega_{\rm max}={\cal F}_{\rm max}\left(\frac{G M_{\rm max}}{R^3_{\rm max}} \right)^{1/2}.
\label{Omega-max-1}
\end{equation}
The factor ${\cal F}_{max}$ depends on the various approximations.
\indent For the numerical integration of the equilibrium equations, we used the public RNS code~\cite{rns} by Stergioulas and Friedman~\cite{Stergioulas-1995} (This code is based on the method developed by Komatsu, Eriguchi and Hachisu~\cite{Komatsu-1989} and modifications introduced by Cook, Shapiro and Teukolsky~\cite{Cook-1994}).

\begin{figure*}
	\centering
	\includegraphics[width=0.49\textwidth]{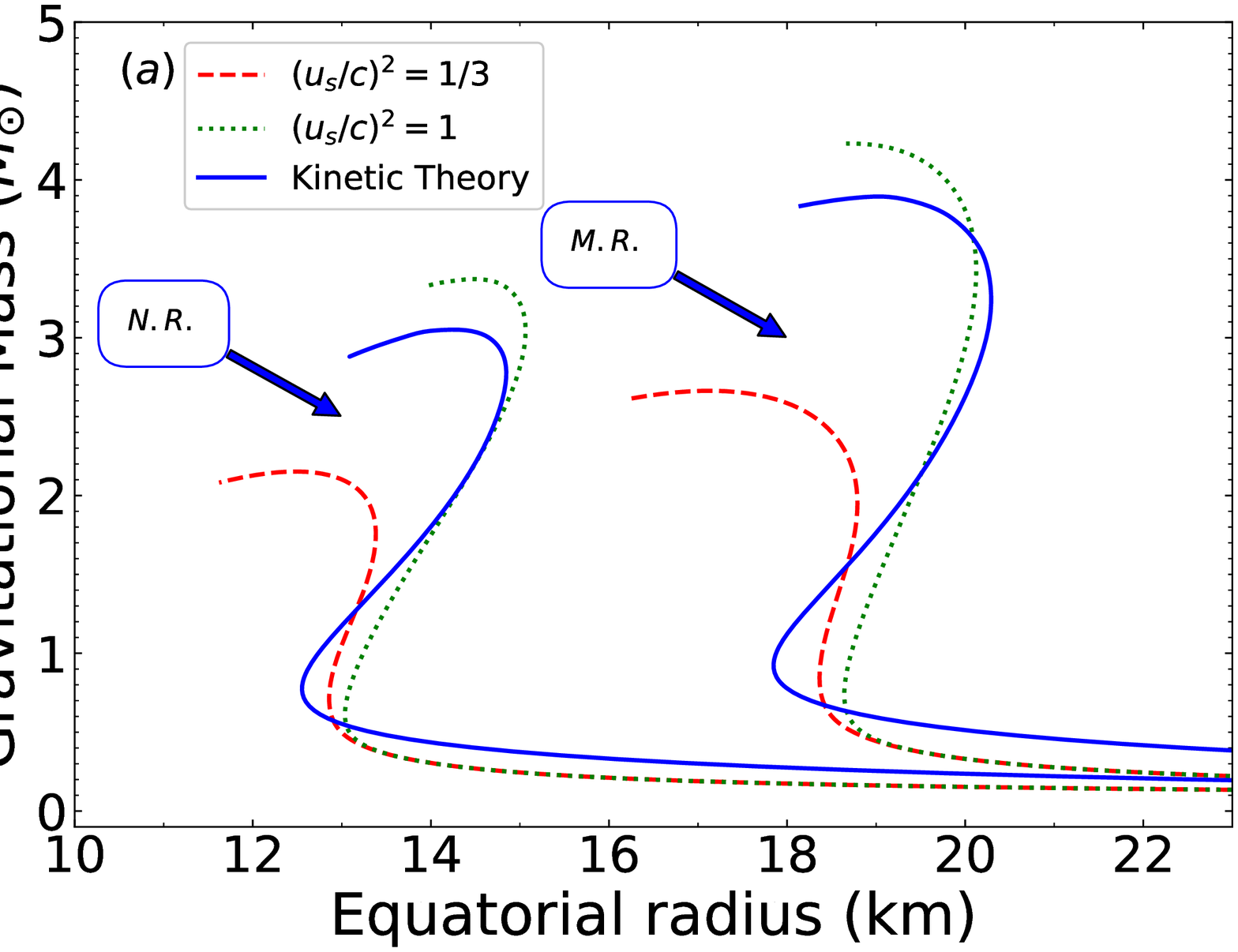}
	~
	\includegraphics[width=0.49\textwidth]{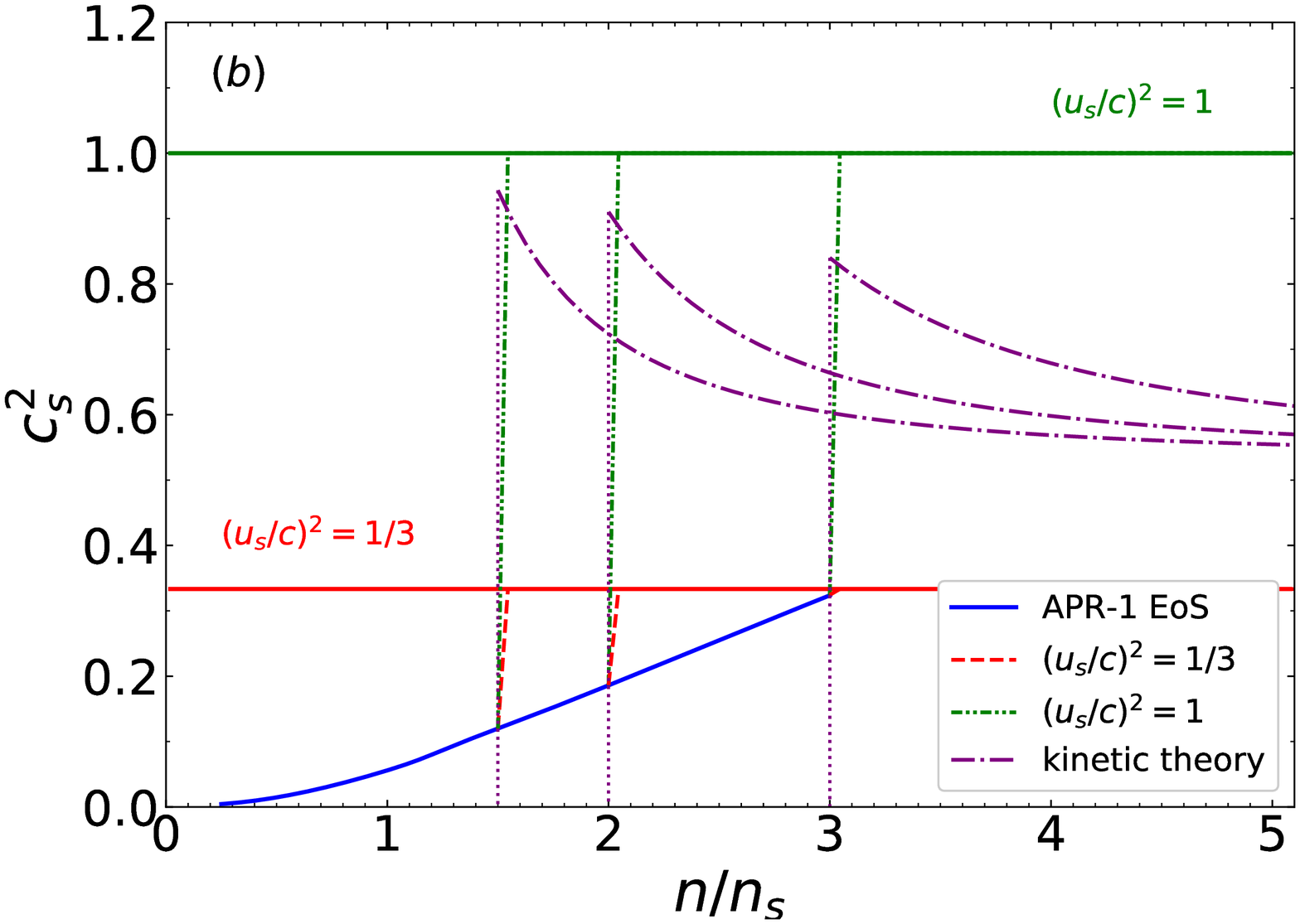}
	\caption{(color online) (a) Mass-radius diagram for various speed of sound bounds and for transition density equal to 1.5$ n_{\rm s}$. Both nonrotating (N.R.) and maximally rotating (M.R.) configurations are presented and noted in the figure. The $c/\sqrt{3}$ bound is presented with the dashed lines, the $c$ bound with the dotted lines and the one from relativistic kinetic theory with the solid lines. (b) Dependence of the speed of sound on the density for three cases of transition density ($n_{\rm tr}/n_{\rm s}=1.5, 2, 3$) and for the various speed of sound bounds. The APR-1 EoS is presented with the blue solid line, the $c/\sqrt{3}$ bound with the lower red horizontal line and the $c$ bound with the upper green horizontal line. The matching of the APR-1 with (i) the $c/\sqrt{3}$ bound is presented with the red dashed lines, (ii) the $c$ bound is presented with the green dash-dot-dotted lines and (iii) the bound from relativistic kinetic theory is presented with the purple dash-dotted lines. The $c/\sqrt{3}$ and $c$ bounds are also indicated. The vertical dotted lines at $1.5n_{\rm s}$, $2n_{\rm s}$ and $3n_{\rm s}$ are guides for the eye.}
	\label{fig:mass_radius}
\end{figure*}

\section{Constraints on the bulk properties} \label{sec:4}
The EoSs that we used are based on the EoS produced by the data from Akmal {\it et al.}~\cite{Akmal-1998} and the MDI model (for more details see Ref.~\cite{Koliogiannis-2019}) and, specifically, the APR-1 EoS. In particular, in Fig.~\ref{fig:mass_radius}(a) we present the gravitational mass dependence on the corresponding equatorial radius for the APR-1 EoS with the various speed of sound bounds and the transition density at $1.5n_{\rm s}$ for both the nonrotating and maximally rotating cases. In Fig.~\ref{fig:mass_radius}(b) we present the speed of sound dependence on the density for three cases of the transition density ($n_{\rm tr}/n_{\rm s}=1.5, 2, 3$) and for the various speed of sound bounds. Obviously, the discontinuity in the speed of sound in the cases of the $c$ and $c/\sqrt{3}$ bounds are treated effectually   with the help of the approximation in Eq.~\eqref{speed-matc-1}. The corresponding discontinuity in the case of the kinetic theory is also exhibited which in our calculations is taken into account without any approximation.

\subsection{Gravitational mass and radius}\label{sec:4a}
The gravitational mass and the corresponding radius (equatorial and polar) of a neutron star can take different values depending both on the assumption of the speed of sound bound and the transition density. We studied the dependence of the gravitational mass and the corresponding radius for the three bounds and also for different values of the transition density, up to 5$n_{\rm s}$.

\begin{figure}[H]
	\includegraphics[width=0.5\textwidth]{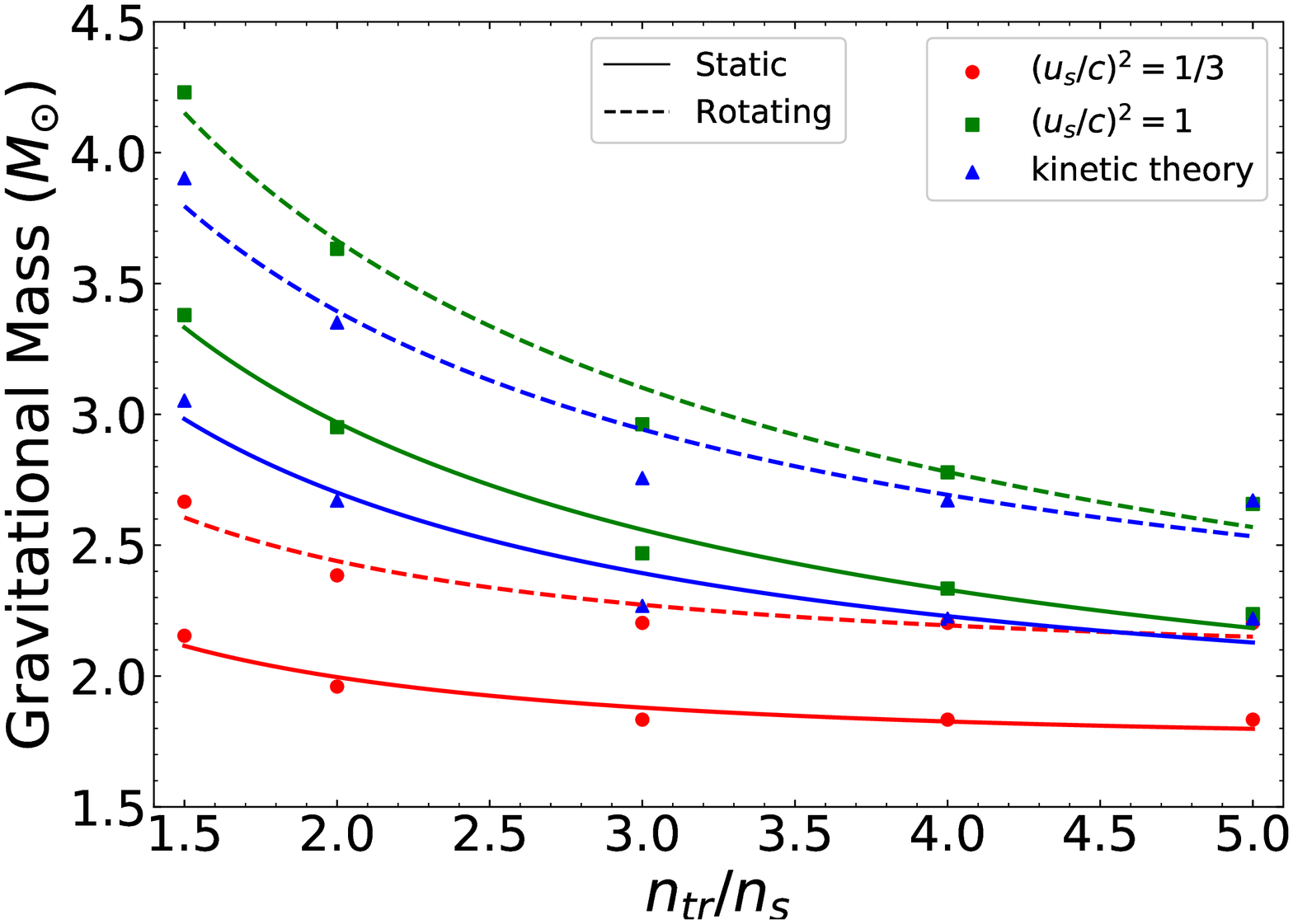}
	\caption{(color online) Dependence of the gravitational mass on the transition density at the maximum mass configuration for the various speed of sound bounds. The data at the maximum mass configuration are presented with circles for the $v_{\rm s}/c=1/\sqrt{3}$ bound, squares for the $v_{\rm s}/c=1$ bound and triangles for the one from relativistic kinetic theory. The fits for the nonrotating and maximally rotating configuration are presented with the solid and dashed lines, respectively.}
	\label{fig:mass_ntr}
\end{figure}

\begin{figure*}
	\centering
	\includegraphics[width=0.49\textwidth]{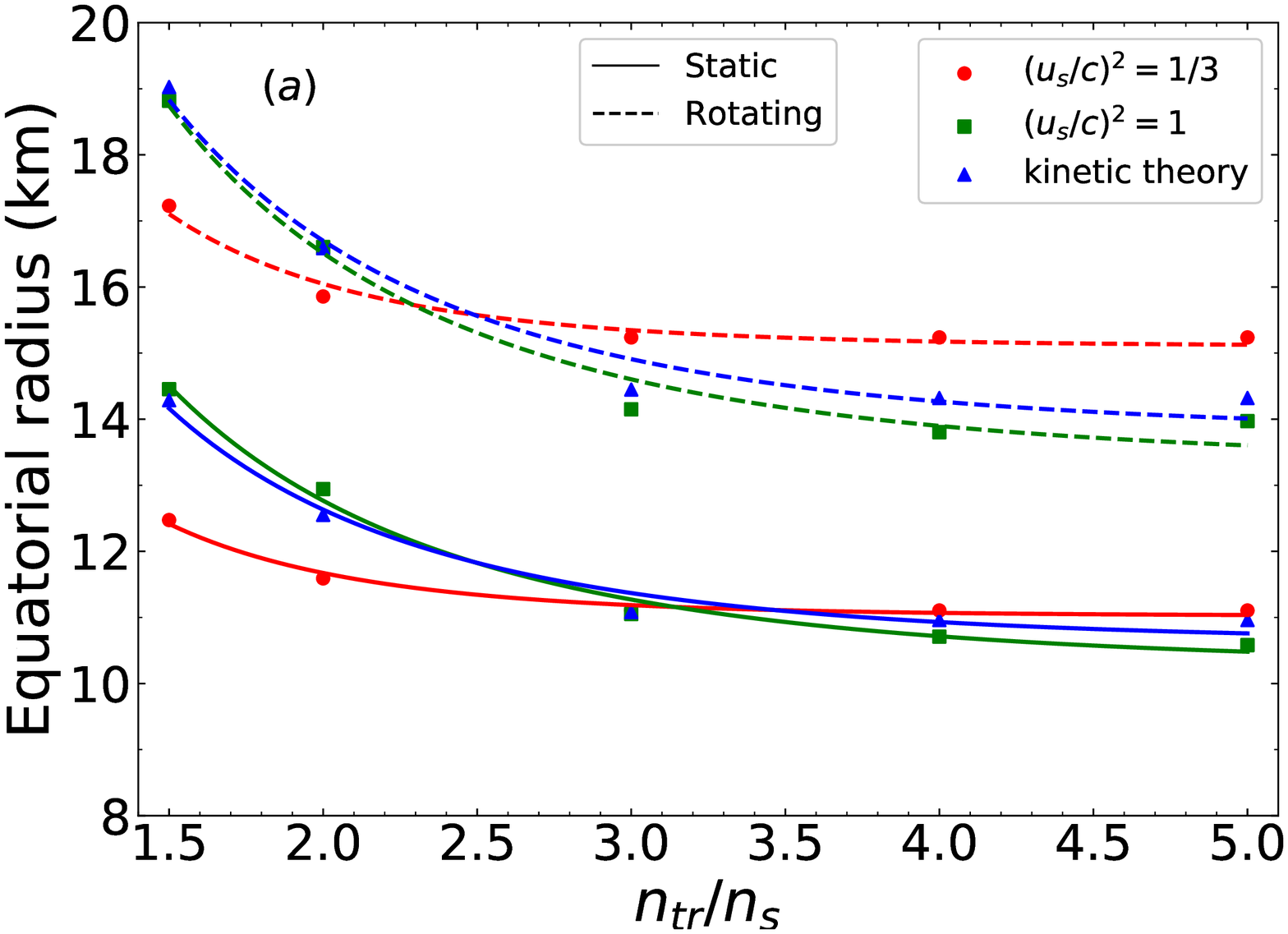}
	~
	\includegraphics[width=0.49\textwidth]{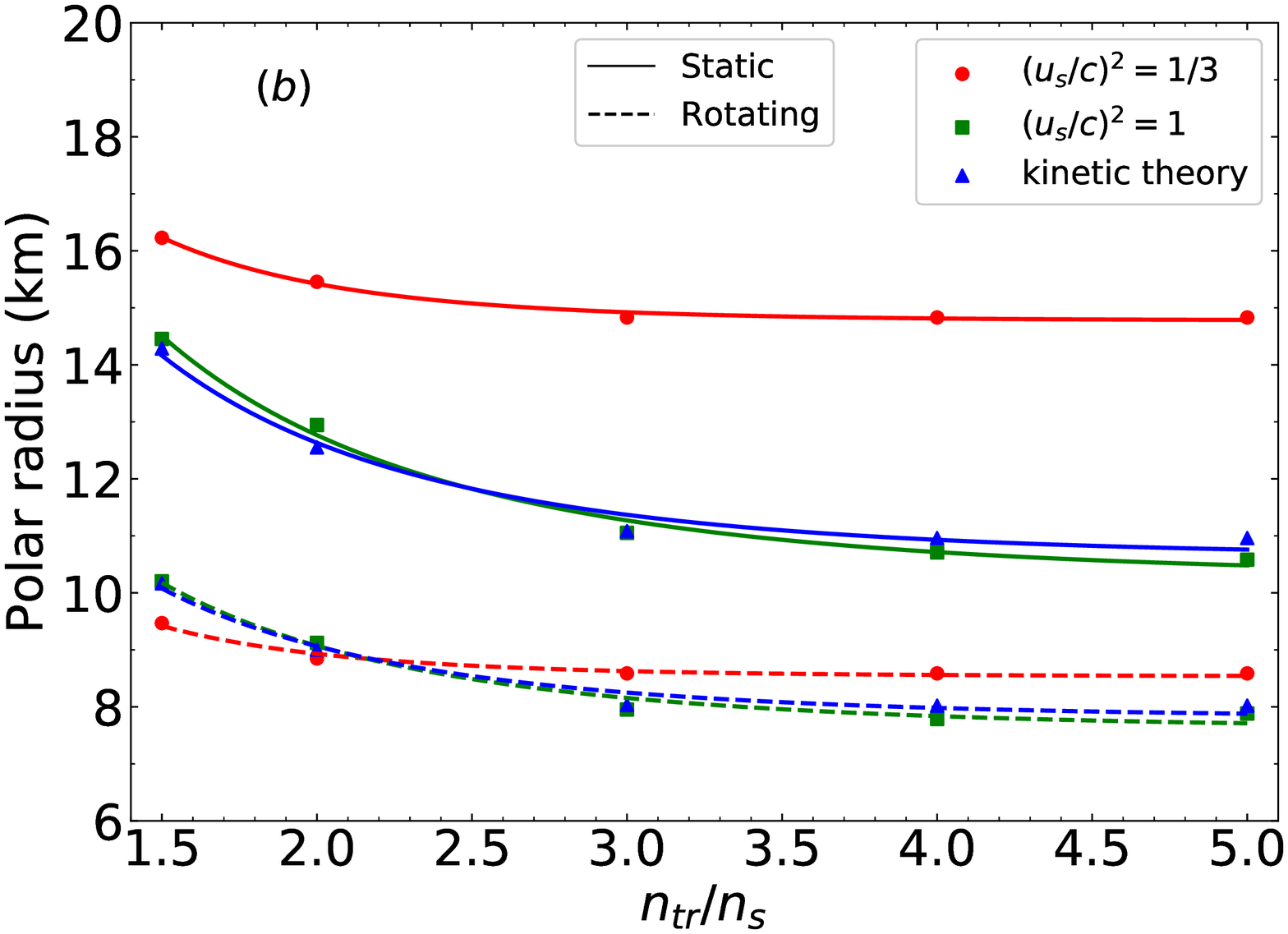}
	\caption{(color online) Dependence of (a) the equatorial and (b) the polar radius on the transition density at the maximum mass configuration for the various speed of sound bounds. The data at the maximum mass configuration are presented with circles for the $v_{\rm s}/c=1/\sqrt{3}$ bound, squares for the $v_{\rm s}/c=1$ bound and triangles for the one from relativistic kinetic theory. The fits for the nonrotating and maximally rotating configuration are presented with the solid and dashed lines, respectively.}
	\label{fig:radius-ntr}
\end{figure*}

\indent In the case of the gravitational mass, as Fig.~\ref{fig:mass_ntr} shows, both in nonrotating and maximally rotating configurations, a reduction on the gravitational mass along the transition density occurs until it reaches a constant value. The dependence of the gravitational mass on the transition density can be described accurately by the formula 

\begin{equation}
	\frac{M_{\rm max}}{M_{\odot}} = \alpha_{1} \coth\left[\alpha_{2}\left(\frac{n_{\rm tr}}{n_{\rm s}}\right)^{1/2}\right],
	\label{eq:mass_n}
\end{equation}
where $\alpha_{1}$ and $\alpha_{2}$ are shown in Table~\ref{tab:2}. For reasons of completeness, we also studied the approximation within Ref.~\cite{Kalogera-1996} for the nonrotating configuration. In fact, this was the study of Hartle~\cite{Hartle-1978}, where he proved the analytical solution which is shown below.
This is valid only in region [1.5-3]$n_{s}$ and it is given through the expression
\begin{equation}
	\frac{M_{\rm max}}{M_{\odot}} = k\sqrt{\frac{1}{n_{\rm tr}/n_{\rm s}}} \approx \frac{\alpha_{1}}{\alpha_{2}}\sqrt{\frac{1}{n_{\rm tr}/n_{\rm s}}},
	\label{eq:mass_n_1}
\end{equation}
\noindent where the coefficient $k$ takes the values 3.319, 5.165 and 4.765 for the bounds $c/\sqrt{3}$, $c$ and the one from relativistic kinetic theory, respectively. It should be noted that Eq.~\eqref{eq:mass_n} for low values of density leads to Eq.~\eqref{eq:mass_n_1}, since the coefficient $k$ is approximately equal to $\alpha_{1}/\alpha_{2}$. \\
\indent Figure~\ref{fig:radius-ntr} displays the equatorial and polar radius as a function of the transition density. While similar dependence with the gravitational mass is presented, the bound $c/\sqrt{3}$ differs. In particular, after 3$n_{s}$ the $c/\sqrt{3}$ case leads to higher values of the radius than the other two bounds. The dependence of the equatorial radius on the transition density can be described accurately by the formula
\begin{equation}
	\frac{R_{\rm max}}{\rm km} = \alpha_{3} \coth\left[\alpha_{4}\left(\frac{n_{\rm tr}}{n_{\rm s}}\right)^{0.8}\right],
	\label{eq:radius_n}
\end{equation}
where $\alpha_{3}$ and $\alpha_{4}$ are shown in Table~\ref{tab:2}. Its specific meaning could be lying under the fact that the bound $c/\sqrt{3}$ can lead to higher deformed models of neutron stars.

\subsection{Mass-shedding angular velocity}\label{sec:4b}
\begin{figure}[H]
	\includegraphics[width=0.5\textwidth]{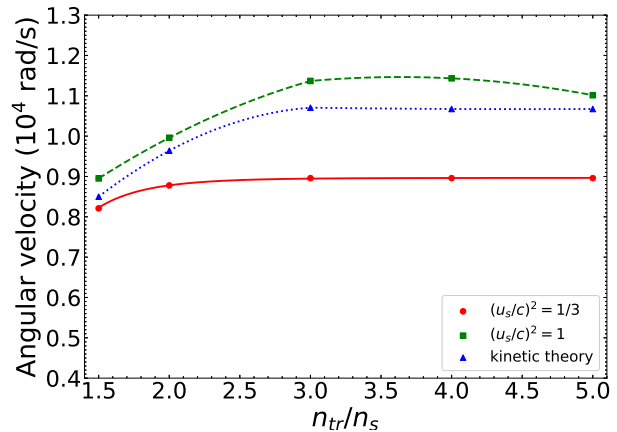}
	\caption{(color online) Dependence of the angular velocity on the transition density at the maximum mass configuration for the various speed of sound bounds. The data at the maximum mass configuration and the fits are presented with the circles and the solid line for the $v_{\rm s}/c=1/\sqrt{3}$ bound, the squares and the dashed line for the $v_{\rm s}/c=1$ bound and the triangles and the dotted line for the one from relativistic kinetic theory.}
	\label{fig:omega_n}
\end{figure}

Figure ~\ref{fig:omega_n} displays the dependence of the Keplerian angular velocity on the transition density. Angular velocity at the maximum mass configuration for the maximally rotating case is an increasing function of the transition density. This effect remains valid only for the bound $c/\sqrt{3}$. In the rest of the cases, it reaches a maximum at $3n_{s}$ and then decreases along the transition density.

\begin{figure*}
	\centering
	\includegraphics[width=0.49\textwidth]{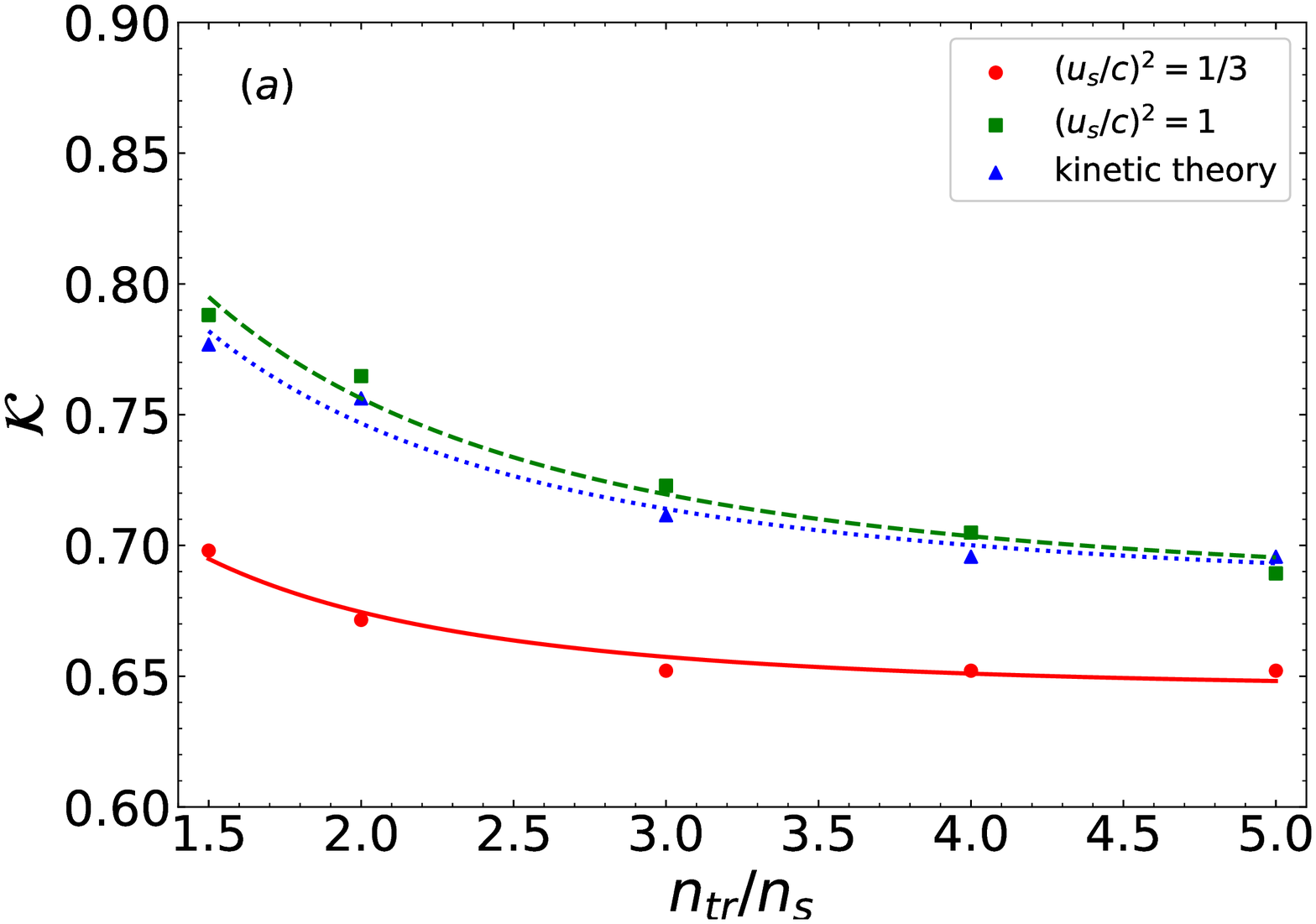}
	~
	\includegraphics[width=0.49\textwidth]{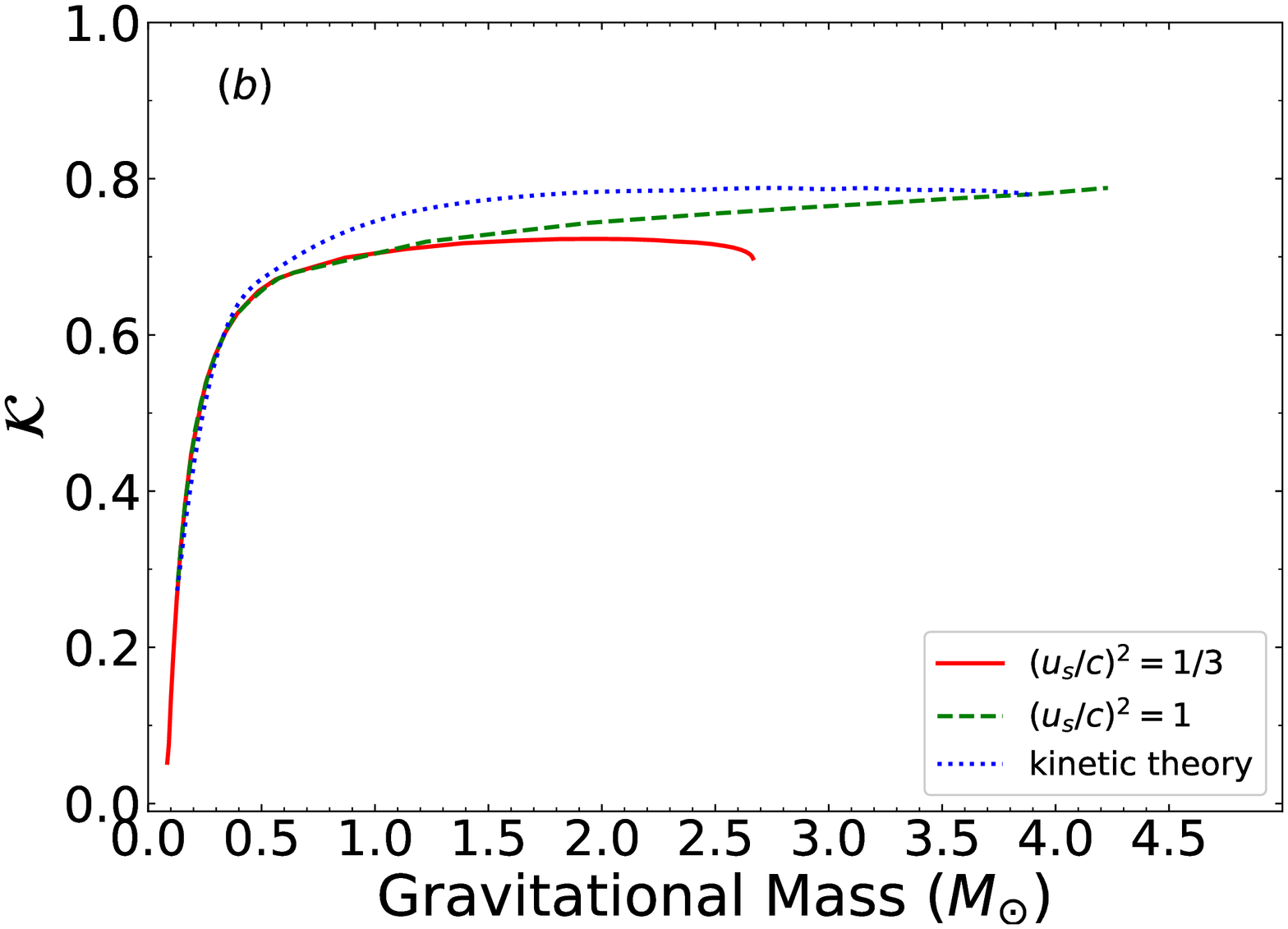}
	\caption{(color online) Dependence of the Kerr parameter ($\mathcal{K}$) on (a) the transition density at the maximum mass configuration and (b) the gravitational mass at $n_{\rm tr} = 1.5n_{\rm s}$ for the various speed of sound bounds. The data at the maximum mass configuration and the fits are presented with the circles and the solid line for the $v_{\rm s}/c=1/\sqrt{3}$ bound, the squares and the dashed line for the $v_{\rm s}/c=1$ bound, and the triangles and the dotted line for the one from relativistic kinetic theory.}
	\label{fig:Kerr_ntr}
\end{figure*}

\indent From Fig.~\ref{fig:omega_n}, it is clear that the maximum angular velocity at the Keplerian sequence is not presented with the max-stiff EoS (which corresponds to $n_{\rm tr}=1.5n_{\rm s}$), but with an EoS with the transition density equals to $3n_{\rm s}$. This effect is probably a direct consequence of the relation that exists between the Keplerian angular velocity and the maximum neutron star mass, as well as the corresponding equatorial radius.

\subsection{Kerr parameter}\label{sec:4c}
The Kerr parameter is important for neutrons stars and in general, compact objects. To be more specific, it can lead to possible limits for the compactness on neutron stars and second, can be a criterion for determining the final fate of the collapse of a rotating compact star~\cite{Lo-2011,Koliogiannis-2019}. The relation which describes the Kerr parameter~\cite{Koliogiannis-2019} is
\begin{equation}
	\mathcal{K} = \frac{cJ}{GM^{2}}.
\end{equation}
\indent The Kerr parameter, as Fig.~\ref{fig:Kerr_ntr}(a) shows, follows the mass-density relation as a decreasing function of the transition density. The dependence of the Kerr parameter on the transition density can be described accurately by the formula
\begin{equation}
	\mathcal{K_{\rm max}} = \alpha_{7} \coth\left[\alpha_{8}\left(\frac{n_{\rm tr}}{n_{\rm s}}\right)^{1/2}\right],
	\label{eq:kerr_n}
\end{equation}
where $\alpha_{7}$ and $\alpha_{8}$ are shown in Table~\ref{tab:2}.\\
\indent From Fig.~\ref{fig:Kerr_ntr}(b), the dependence of this parameter on the gravitational mass for the max-stiff EoS in each case can lead us to possible constraints on the maximum neutron star mass. Both of these figures may help to constrain the EoS, by constraining the Kerr parameter and the possible maximum mass of a neutron star almost simultaneously.

\begin{table*}
	\squeezetable
	\caption{Coefficients of Eqs.~\eqref{eq:mass_n},~\eqref{eq:radius_n},~\eqref{eq:kerr_n} and ~\eqref{eq:moi_n} for the three speed of sound bounds. The abbreviation ``n.r." corresponds to the nonrotating configuration and the ``m.r." to the maximally rotating one.}
	\begin{ruledtabular}
		\begin{tabular}{ccccccccccccccccc}
			\multirow{2}{*}{Speed of sound bounds} & \multicolumn{2}{c}{$\alpha_{1}$} & \multicolumn{2}{c}{$\alpha_{2}$} & \multicolumn{2}{c}{$\alpha_{3}$} & \multicolumn{2}{c}{$\alpha_{4}$} & \multicolumn{2}{c}{$\alpha_{5}$} & \multicolumn{2}{c}{$\alpha_{6}$} & \multicolumn{2}{c}{$\alpha_{7}$} & \multicolumn{2}{c}{$\alpha_{8}$}\\
			& n.r. & m.r. & n.r. & m.r. & n.r. & m.r. & n.r. & m.r. & n.r. & m.r. & n.r. & m.r. & n.r. & m.r. & n.r. & m.r.\\
			\hline
			$c$ & 1.665 & 1.689 & 0.448 & 0.352 & 10.280 & 13.352 & 0.639 & 0.645 & -- & 3.259 & -- & 0.133 & -- & 0.683 & -- & 1.053 \\
			
			$c/\sqrt{3}$ & 1.751 & 2.069 & 0.964 & 0.883 & 11.024 & 15.110 & 1.021 & 1.009 & -- & 3.040 & -- & 0.362 & -- & 0.645 & -- & 1.348 \\
			
			kinetic theory & 1.844 & 1.999 & 0.590 & 0.478 & 10.629 & 13.802 & 0.704 & 0.676 & -- & 3.533 & -- & 0.167 & -- & 0.683 & -- & 1.101 \\
		\end{tabular}
	\end{ruledtabular}
	\label{tab:2}
\end{table*}

\subsection{Moment of inertia}\label{sec:4d}
One of the most relevant properties in pulsar analysis is the moment of inertia which can quantify how fast an object can spin with a given angular momentum~\cite{Cipolletta-2015,Koliogiannis-2019}. It is given by the form
\begin{equation}
	I = \frac{J}{\Omega},
\end{equation}
where $J$ is the angular momentum and $\Omega$ the angular velocity of the neutron star.

\begin{figure}[H]
	\includegraphics[width=0.5\textwidth]{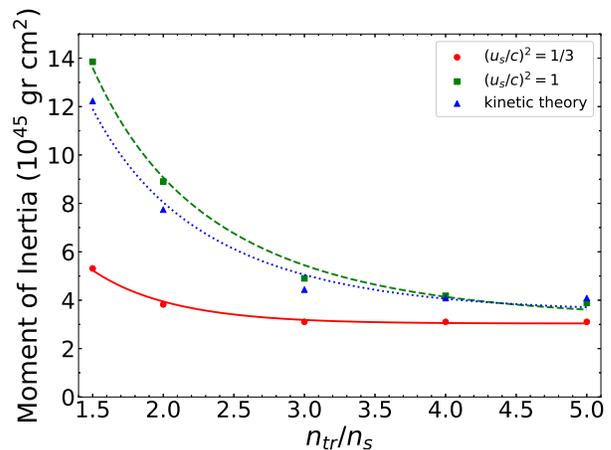}
	\caption{(color online) Dependence of the moment of inertia on the transition density at the maximum mass configuration for the various speed of sound bounds. The data at the maximum mass configuration and the fits are presented with the circles and the solid line for the $v_{\rm s}/c=1/\sqrt{3}$ bound, the squares and the dashed line for the $v_{\rm s}/c=1$ bound and the triangles and the dotted line for the one from relativistic kinetic theory.}
	\label{fig:moi_n}
\end{figure}

\begin{figure*}
	\includegraphics[width=0.9\textwidth]{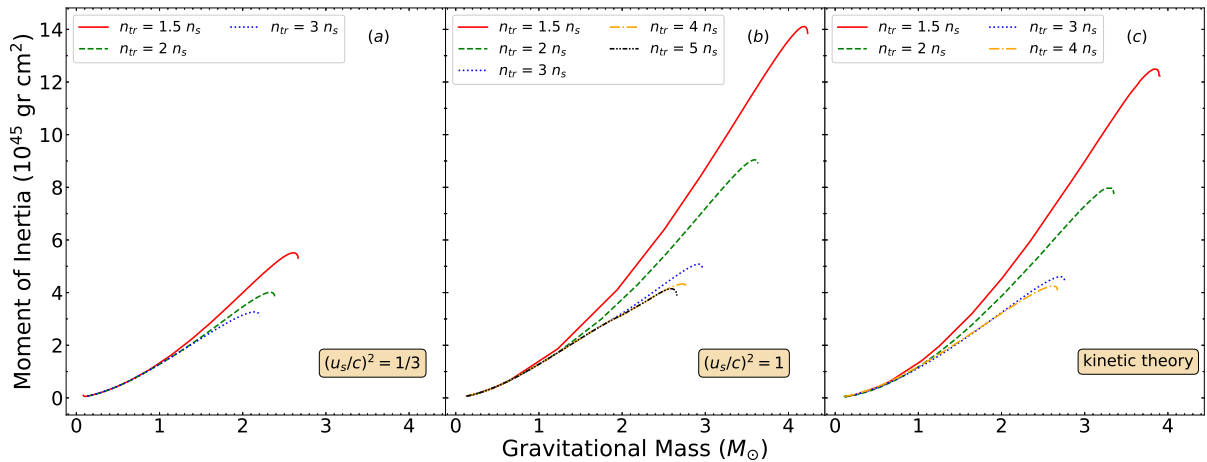}
	\caption{(color online) Dependence of the moment of inertia on the gravitational mass for the bounds (a) $v_{\rm s}/c = 1/\sqrt{3}$, (b) $v_{\rm s}/c = 1$ and (c) the one from relativistic kinetic theory and for the various transitions densities. $n_{\rm tr}=1.5n_{\rm s}$ is presented with the solid lines, $n_{\rm tr}=2n_{\rm s}$ is presented with the dashed lines, $n_{\rm tr}=3n_{\rm s}$ is presented with the dotted lines, $n_{\rm tr}=4n_{\rm s}$ is presented with the dash-dotted lines and $n_{\rm tr}=5n_{\rm s}$ is presented with the dash-dot-dotted lines.}
	\label{fig:moi_m}
\end{figure*}

\indent In Fig.~\ref{fig:moi_n}, we studied the moment of inertia in correlation with the transition density for the various bounds. In this case too, the moment of inertia follows a decreasing trajectory along the transition density until it reaches a constant value. The significant meaning of this property appears with the $v_{\rm s}/c = 1/\sqrt{3}$ bound which leads to much lower values for the moment of inertia than the other two. The dependence of the moment of inertia on the transition density can be described accurately by the formula
\begin{equation}
	I_{\rm max} = \alpha_{5}\coth\left[\alpha_{6} \left(\frac{n_{\rm tr}}{n_{\rm s}}\right)^{3/2}\right] \quad \left(10^{45}\text{ }\rm  gr \text{ } \rm cm^{2}\right),
	\label{eq:moi_n}
\end{equation}
where $\alpha_{5}$ and $\alpha_{6}$ are shown in Table~\ref{tab:2}. For reasons of completeness, we also studied the approximation within Ref.~\cite{Silva_2017} for the maximally rotating configuration. This is valid only in region [1.5-3]$n_{s}$ and it is given through
\begin{equation}
I_{\rm max}= k_{2}\left(\frac{1}{n_{\rm tr}/n_{\rm s}}\right)^{k_{3}} \quad \left(10^{45}\text{ }\rm  gr \text{ } \rm cm^{2}\right),
\label{eq:moi_n_1}
\end{equation}      
where $k_{2}$ and $k_{3}$ are shown in Table~\ref{tab:3}.\\
\indent In addition, we studied the dependence of the moment of inertia on the gravitational mass for the various bounds. In Fig.~\ref{fig:moi_m}, we can see that, in all cases, the minimum transition density (max-stiff EoS) leads to the maximum possible moment of inertia.\\

\begin{table}[H]
	\squeezetable
	\caption{Coefficients of Eq.~\eqref{eq:moi_n_1} for the three speed of sound bounds.}
	\begin{ruledtabular}
		\begin{tabular}{ccc}
			Speed of sound bounds & $k_{2}$ & $k_{3}$ \\
			\hline
			
			$c$ & 24.973 & 1.471 \\
			$c/\sqrt{3}$ & 8.476 & 1.150 \\
			kinetic theory & 22.337 & 1.498 \\
		\end{tabular}
	\end{ruledtabular}
	\label{tab:3}
\end{table}

\subsection{Constant rest mass sequences}\label{sec:4e}
\begin{figure}[H]
	\includegraphics[width=0.46\textwidth]{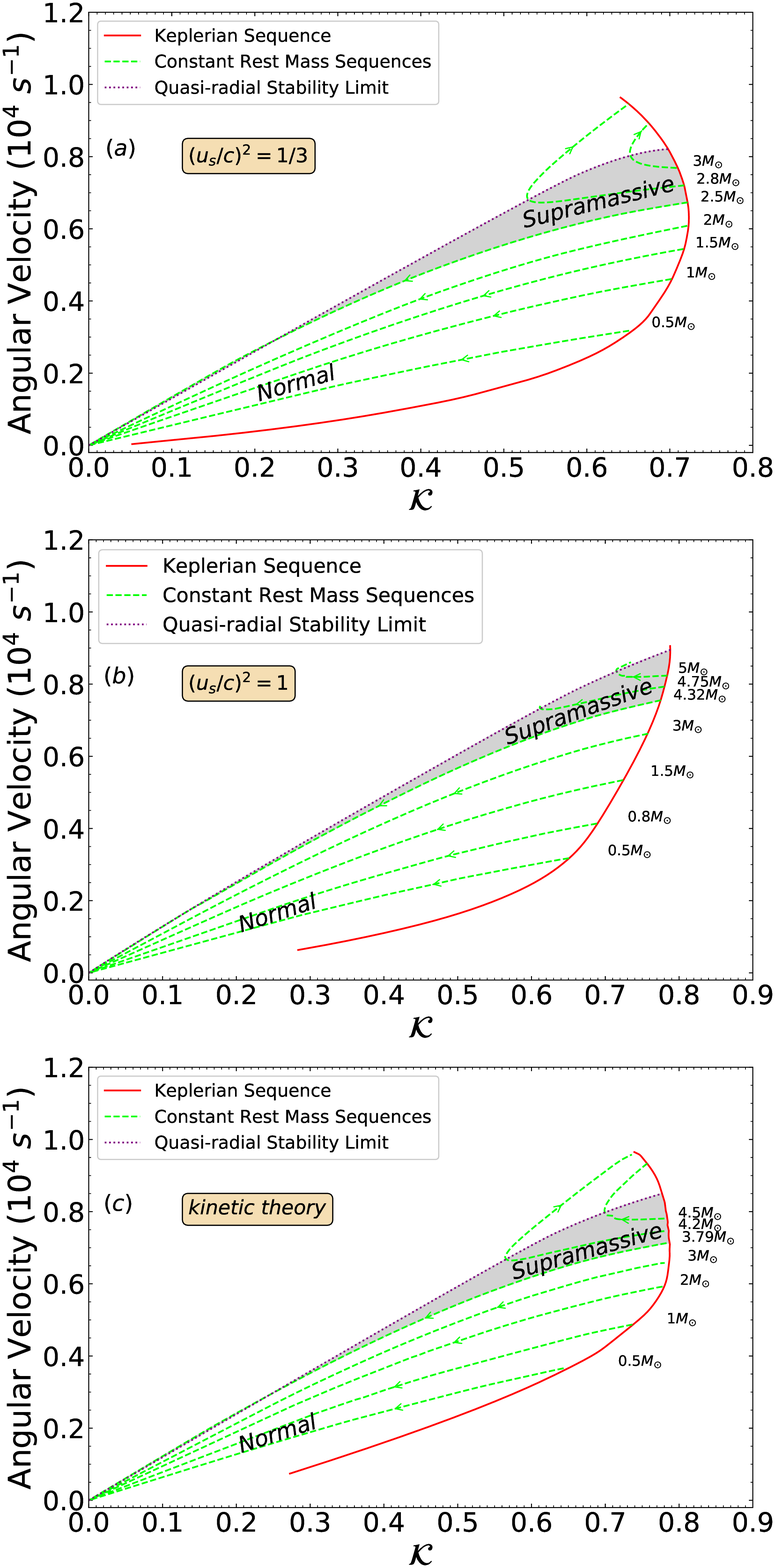}
	\caption{(color online) Normal and supramassive rest mass sequences as the dependence of the angular velocity on the Kerr parameter for the speed of sound bounds (a) $v_{\rm s}/c=1/\sqrt{3}$, (b) $v_{\rm s}/c=1$ and (c) the one from relativistic kinetic theory at $n_{\rm tr}=1.5n_{\rm s}$. The red curves represent the Keplerian sequence. The constant rest mass sequences are presented with the green dashed lines and the quasi-radial stability limit with the purple dotted lines.}
	\label{fig:omega_k}
\end{figure}

In order to study the time evolution of a neutron star in correlation with the transition density and the various speed of sound bounds, we present in Fig.~\ref{fig:omega_k} the normal and supramassive rest mass sequences as the dependence of the angular velocity on the Kerr parameter for various values of the rest mass at $n_{\rm tr}=1.5n_{\rm s}$. To be more specific, these sequences describe the time evolution of a neutron star (with a fixed rest mass) created by spinning with its Keplerian velocity. In the case of normal ones, neutron stars eventually losing their angular momentum (for various reasons) and becoming nonrotating as they approach to a stable configuration. On the other

\begin{figure}[H]
	\includegraphics[width=0.5\textwidth]{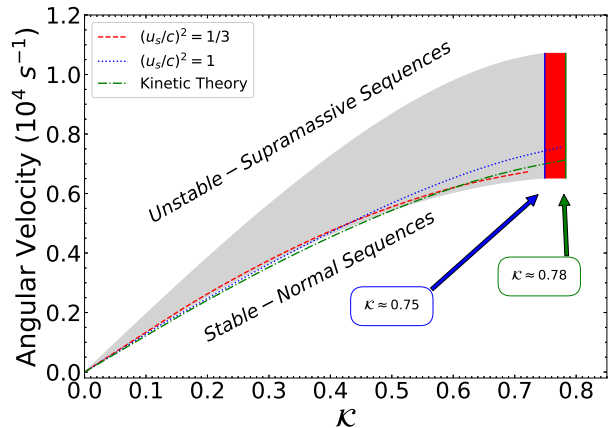}
	\caption{(color online) Last stable rest mass sequence as the dependence of the angular velocity on the Kerr parameter for the various speed of sound bounds at $n_{\rm tr}=1.5n_{\rm s}$. Supramassive and normal areas are noted. The dashed line represents the $v_{\rm s}/c=1/\sqrt{3}$ bound, the dotted line represents the $v_{\rm s}/c=1$ bound and the dashed-dotted line represents the one from relativistic kinetic theory. The gray area and Kerr limit 0.75 from Ref.~\cite{Koliogiannis-2019} and the red area from the present work are also presented. The maximum value at each case is also noted.}
	\label{fig:lsrms}
\end{figure}

\noindent hand, during the supramassive sequence, neutron stars never approach the stable nonrotating case and their final fate is to collapse to a black hole. Obviously, the possible upper bounds on the speed of sound affect the time evolution of maximally rotating neutron stars. The latter depends mainly on the value of the rest mass at which the star is created.\\
\indent As a follow up to Fig.~\ref{fig:omega_k}, we plot in Fig.~\ref{fig:lsrms} the last stable rest mass sequence (LSRMS) for the various bounds at $n_{\rm tr}=1.5n_{\rm s}$. This specific sequence is the one that corresponds to the maximum mass configuration at the nonrotating model and defines the upper limit to the stable region. Moreover, in the same figure we plot the results concerning the LSRMS found by employing a large number of realistic EoSs~\cite{Koliogiannis-2019} [shadowed (gray) region]. It is remarkable that the three bounds define the lower limit on the LSRMS. This means that the effect of the bounds is to lower the stable region of the time evolution concerning the angular velocity. However, in this study, there is no striking signature of the bounds under consideration since all of them lead to  almost similar trends.

\subsection{Minimum rotational period}\label{sec:4f}
\begin{figure}[H]
	\includegraphics[width=0.5\textwidth]{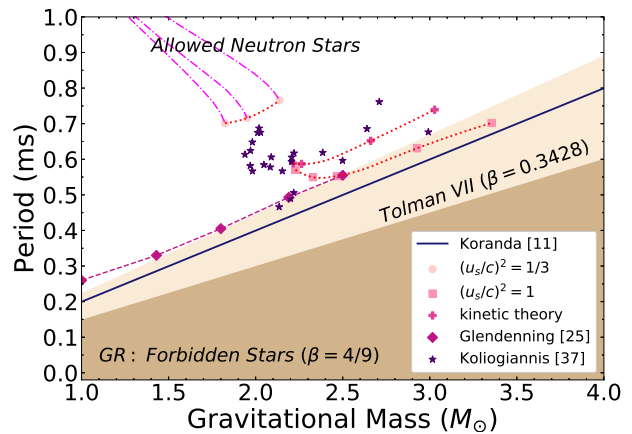}
	\caption{(color online) Minimum rotational period of rotating neutron stars as a function of the mass of the maximum mass spherical star allowed by the EoS of the stellar matter. The circles represent the $v_{\rm s}/c = 1/\sqrt{3}$ bound, the squares represent the $v_{\rm s}/c = 1$ bound and the crosses represent the one from relativistic kinetic theory. The data from Glendenning~\cite{Glendenning-1992} are also presented with diamonds. The blue curve is adapted from Koranda {\it et al.}~\cite{Koranda-1997} for comparison. The data for 23 hadronic realistic EoS~\cite{Koliogiannis-2019} are also presented with stars. The region forbidden from the structure of General Relativity (GR) for $\beta =4/9$ and the one from Tolman VII with $\beta =0.3428$ are presented as guides.} 
	\label{fig:period}
\end{figure}

\indent For a given EoS the maximum gravitational mass in the sequence of gravitationally bound neutron stars has the minimum Keplerian period. Therefore, in order to provide useful constraints on the EoS, we present in Fig.~\ref{fig:period} the dependence of the minimum rotational period as a function of the corresponding maximum gravitational mass of the nonrotating configuration for the various speed of sound bounds and for the max-stiff EoS in each case.

\begin{table*}
	\squeezetable
	\caption{The values of the  central density $\rho_{\rm c}={\cal E}_{\rm c}/c^2$ (in units of $10^{15}$ $\rm gr$ $\rm cm^{-3}$) which correspond to each transition density at the maximum mass configuration for the various speed of sound bounds. The abbreviation ``n.r." corresponds to the nonrotating configuration and the ``m.r." to the maximally rotating one.}
	\begin{ruledtabular}
		\begin{tabular}{ccccccccccc}
			\multirow{2}{*}{Speed of sound bounds} & \multicolumn{2}{c}{$n_{\rm tr}=1.5n_{\rm s}$} & \multicolumn{2}{c}{$n_{\rm tr}=2n_{\rm s}$} & \multicolumn{2}{c}{$n_{\rm tr}=3n_{\rm s}$} & \multicolumn{2}{c}{$n_{\rm tr}=4n_{\rm s}$} & \multicolumn{2}{c}{$n_{\rm tr}=5n_{\rm s}$} \\
			& n.r. & m.r. & n.r. & m.r. & n.r. & m.r. & n.r. & m.r. & n.r. & m.r.\\
			\hline
			$c$ & 1.239 & 0.993 & 1.463 & 1.309 & 2.276 & 2.038 & 2.405 & 2.153 & 2.542 & 2.153 \\
			
			$c/\sqrt{3}$ & 1.726 & 1.369 & 2.038 & 1.726 & 2.276 & 1.928 & -- & -- & -- & -- \\
			
			kinetic theory & 1.228 & 1.000 & 1.689 & 1.400 & 2.384 & 2.000 & 2.384 & 2.095 & -- & -- \\
		\end{tabular}
	\end{ruledtabular}
	\label{tab:4}
\end{table*}

\indent In addition, just for comparison, the results of the previous work of Glendenning~\cite{Glendenning-1992} as well as Koranda {\it et al.}~\cite{Koranda-1997} have been included. Moreover, the results corresponding to 23 hadronic EoSs from Koliogiannis and Moustakidis~\cite{Koliogiannis-2019} are presented. Finally, the two forbidden regions provided by the maximum value of the compactness parameter, $\beta=GM/Rc^2$, are included. In the first one, the value is $\beta=4/9$ (the maximum value of compactness  allowed by General Relativity)~\cite{Glendenning-2000} and in the second one is $\beta=0.3428$ which is the maximum possible value of the Tolman-VII analytical solution of the Tolman-Oppenheimer-Volkoff [(TOV) Einstein's equations for a nonrotating spherical symmetric object] that leads to a stable configuration. It was found recently that this value is a universal upper limit also including realistic EoSs~\cite{Koliogiannis-2019-b}. The above two regions have been found by combining the expression
\begin{equation}
P_{\rm min}={\cal F} \left(\frac{M_{\odot}}{M^{\rm st}_{\rm max}}\right)^{1/2}\left(\frac{R^{\rm st}_{\rm max}}{10 \rm km}\right)^{3/2} \quad ({ms}),
\label{Pmin-1}
\end{equation}
which provided very recently in Ref.~\cite{Koliogiannis-2019} and relates the minimum period of a maximally rotating neutron star with the mass and radius of the nonrotating case and the limited values of the compactness parameter, $\beta_{max}$. In this case we have found the expression
\begin{equation}
P_{\rm min}=0.79\left(\frac{0.1473}{\beta_{\rm max}}\right)^{3/2}\left(\frac{M^{\rm st}_{\rm max}}{M{\odot}}\right) \quad ({ms}),
\label{Pmin-2}
\end{equation}
which employed in Fig.~\ref{fig:period}.\\
\indent It is remarkable that the approximation of Koranda {\it et al.}~\cite{Koranda-1997} correctly provides the lower limit of the $P_{\rm min}$ compared to the results of the realistic EoSs. In addition, the approximation given by Glendenning~\cite{Glendenning-1992}, even its simplicity, is very close to the previous one. As far as concerning the Tolman-VII solution, its prediction close to the Glendenning's approximation confirms its reliability. However, the most striking feature is the constraints imposed by the bound $v_{\rm s}/c=1/\sqrt{3}$. Obviously, the use of this bound leads to a significant increase in the minimum period  (almost twice the time of the causality limit $v_s/c=1$) and consequently limits the allowed region. In fact, the allowed region is restricted dramatically by using this bound, also excluding the prediction of all realistic EoSs.\\
\indent In the same figure, and for the bound $v_{\rm s}/c=1/\sqrt{3}$, the sequences for each case (which correspond to various values of the critical density and consequently different M-R diagrams) are included. This line helps to identify the minimum period (and the allowed period) for each case and for various values of the mass. As expected, the bound  $v_{\rm s}/c=1/\sqrt{3}$ provides strong constraints on the maximum mass (both on nonrotating and maximally rotating neutron stars), and also on the minimum allowed rotating period.

\begin{table*}
	\squeezetable
	\caption{Comparison between the two approaches presented in this study (see Section~\ref{sec:4h}) for the bulk properties of neutron stars at the maximum mass configuration for $n_{\rm tr} = 1.5 n_{\rm s}$. The abbreviation ``c." corresponds to the continuity method and ``dc." to the discontinuity one.}
	\begin{ruledtabular}
		\begin{tabular}{ccccccccccc}
			\multirow{2}{*}{Speed of sound bounds} & \multicolumn{2}{c}{$M$ $(M_{\odot})$} & \multicolumn{2}{c}{$R$ $(\rm km)$} & \multicolumn{2}{c}{$\Omega$ $(10^{4}$ $\rm s^{-1})$} & \multicolumn{2}{c}{$\mathcal{K}$} & \multicolumn{2}{c}{$I$ $(10^{45}$ $\rm gr$ $\rm cm^{2})$} \\
			& c. & dc. & c. & dc. & c. & dc. & c. & dc. & c. & dc. \\
			\hline
			$c$ & 4.231 & 4.229 & 18.812 & 18.800 & 0.895 & 0.896 & 0.788 & 0.786 & 13.850 & 13.799 \\
			
			$c/\sqrt{3}$ & 2.666 & 2.673 & 17.228 & 17.128 & 0.821 & 0.829 & 0.698 & 0.702 & 5.312 & 5.317 \\
		\end{tabular}
	\end{ruledtabular}
	\label{tab:5}
\end{table*}

\subsection{Central density} \label{sec:4g}
In Table~\ref{tab:4} we present the central density of the star that corresponds to each transition density according to Fig.~\ref{fig:mass_ntr} for both the nonrotating and maximally rotating cases. Actually, the central energy density plays an important role in pulsar analysis~\cite{Koliogiannis-2019}. Moving from the maximally rotating to the non-rotating pulsar, the central density increases, and it is conjectured that the highly compressible quark matter will replace the existing nuclear matter. This effect is directly connected to the reduction of the moment of inertia. Henceforth, the central density can inform us on the appearance of a phase transition in its interior. The latter can lead to the important back-bending phenomenon in pulsars~\cite{Glendenning-2000}.

\subsection{Results from the different approaches} \label{sec:4h}
Two different approaches for the construction of the EoS in correlation with the transition density are studied in this paper. In the first case we studied the method where discontinuities are presented in the EoS while in the second one, we studied the method where continuity exhibits. From the results that are shown in Table~\ref{tab:5}, it is obvious that the two methods converge, especially at the maximum mass configuration, and as a consequence, the effects of the discontinuity are negligible.

\section{Discussion and Conclusions} \label{sec:5}
Sequences of both nonrotating and rotating neutron stars with different speed of sound bounds and transition densities have been studied. In this paper, we have studied the bulk properties of maximally rotating neutron stars in correlation with the transition density and the gravitational mass. In particular, we have calculated their gravitational mass and radius, angular velocity, dimensionless spin parameter and moment of inertia. Relations between these properties and the transition density, as well as the gravitational mass for the various bounds, have been found and shown in the corresponding figures.\\
\indent In particular, the dependence of the gravitational mass, as well as the radius, on the transition density at the maximum mass configuration in the Keplerian sequence for different speed of sound bounds, has been obtained. The gravitational mass and radius are decreasing functions along the transition density for all bounds until they reach a constant value. Relations for these behaviors have been found and shown in the corresponding Sections. The advantage of having these relations falls under the fact that they can describe the full region of the transition density, in contradiction with the old ones, which merely can describe the region [1.5-3]$n_{\rm s}$. Another interesting effect is provided through the $c/\sqrt{3}$ bound, where the radius exceeds the other two bounds after $3n_{\rm s}$. This effect can lead to possible insights for the deformation of the star and also to constrain their radius, as it is one of the open problems in modern astrophysics.\\
\indent Afterward, we have studied the dependence of the angular velocity on the transition density at the maximum mass configuration in the Keplerian sequence for different speed of sound bounds. Although in the case of transition density 1.5$n_{\rm s}$ we have the stiffest EoS, the maximum angular velocity has been achieved at 3$n_{\rm s}$ for each case. To be more specific, in the case of bound $c$ and the one from relativistic kinetic theory, after 3$n_{\rm s}$, a decrease in angular velocity has been observed. On the other hand, in the $c/\sqrt{3}$ case, the angular velocity stabilizes after 3$n_{\rm s}$ in a constant value. The importance of this result must be taken into account with the gravitational mass, as well as the corresponding radius, because Keplerian angular velocity has a complicated dependence on these bulk properties~\cite{Koliogiannis-2019}.\\
\indent In addition, the dependence of the Kerr parameter and moment of inertia on the transition density at the maximum mass configuration in the Keplerian sequence, as well as on the gravitational mass, for different speed of sound bounds have also been obtained.  In the first case, concerning the Kerr parameter, the bound $c$ has led to higher values of the Kerr parameter than the other two bounds. At the most extreme configuration studied in this paper, meaning at 1.5$n_{s}$ and bound $c$, the Kerr parameter has reached a maximum value at around 0.8. The significance of this result for neutron stars  falls under the fact that the gravitational collapse of a uniformly rotating neutron star, constrained to mass-energy and angular momentum conservation, cannot lead to a maximally rotating Kerr black hole~\cite{Koliogiannis-2019}. In the second case, also at the most extreme configuration, the moment of inertia has reached a maximum value at around 14 $\cdot 10^{45}$ $\rm gr$ $\rm cm^{2}$, while in $c/\sqrt{3}$ case the maximum value is at only around 6 $\cdot 10^{45}$ $\rm gr$ $\rm cm^{2}$. The effects of the speed of sound bounds are enhanced for the moment of inertia, where the decrease is up to 2.5 times between the $c$ and $c/\sqrt{3}$ bounds. The insight of this information, because moment of inertia quantifies how fast an object can spin, can lead to possible constraints on the spin frequency on neutron stars.\\
\indent Constant rest mass sequences have also been constructed and studied in order to provide the effects of the speed of sound bounds on the time evolution of neutron stars. As a consequence, the maximum possible rest mass value has been constrained through the different bounds, with the minimum one being $v_{\rm s}/c =1/\sqrt{3}$. In order to provide useful information and constraints on the stable region of neutron stars, we have constructed the LSRMS. From this sequence it is obvious that the normal region of neutron stars is extended downward concerning the angular velocity. The latter has led us to lower spin frequencies on neutron stars.\\
\indent Furthermore, the minimum rotating period of a neutron star as a function of the spherical gravitational mass has been studied. The bound $v_{\rm s}/c = 1/\sqrt{3}$ significantly limits the allowed area of neutron stars, also excluding a majority of realistic EoSs. This bound can provide strong constraints on the maximum gravitational mass, as well as on the minimum rotating period of neutron stars.\\ 
\indent Moreover, two different approaches for the construction of the EoS, one with discontinuity from the transition density and one with continuity, have been studied. These methods converge, as the maximum mass configuration is independent from the low density area of the EoS.\\
\indent Very recent detections of neutron star mergers and the measurements of the corresponding radiated gravitational waves are a powerful tool to study compact objects, mainly including neutron stars and black holes. In particular, these types of observations will be able to provide us with their Keplerian frequency. To be more specific, the remnant formed in the immediate aftermath of the GW170817 merger~\cite{PhysRevLett.119.161101} as well as the recent GW190425 merger~\cite{collaboration2020gw190425}, although believed to have been differentially rotating and not uniformly, contains sufficient angular momentum to be near its mass-shedding limit~\cite{Lattimer-2019}. The observational measurement of the Keplerian frequency as well as of the rest bulk properties (mass, radius, Kerr parameter, moment of inertia, etc.), along with the theoretical predictions, would provide us with strong constraints on the high-density part of the EoS. Moreover, these observations will also help one to check the validity of the proposed upper bounds of the speed of sound in dense nuclear matter (see, for example, the recent Ref.~\cite{Alsing-2018}).

\section*{Acknowledgments}
The authors thank Prof. K. Kokkotas for his constructive comments on the preparation of the manuscript, Prof. N. Stergioulas for providing the RNS code and for the fruitful discussions and Dr. S. Typel for the useful correspondence and discussions. This work was partially supported by the COST action PHAROS (CA16214) and by the DAAD Germany-Greece Grant No. 57340132.

\bibliography{pkoliogiannis.bib}

\end{document}